# Nonvolatile Multilevel States in Multiferroic Tunnel Junctions

Mei Fang[1,2], Sangjian Zhang,[1] Wenchao Zhang,[1] Lu Jiang[3], Eric Vetter[4], Ho Nyung Lee[3], Xiaoshan Xu[5]*, Dali Sun[4]*, and Jian Shen[2]*

[1]*Hunan Key Laboratory of Super Microstructure and Ultrafast Process, School of Physics and Electronics, Central South University, Changsha, Hunan 410083, China*

[2]*State Key Laboratory of Surface Physics and Department of Physics and Collaborative Innovation Center of Advanced Microstructure, Fudan University, Shanghai 200433, China.*

[3]*Materials Science and Technology Division, Oak Ridge National Laboratory, Oak Ridge, Tennessee 37831, USA.*

[4]*Department of Physics, North Carolina State University, Raleigh, North Carolina, 27695, USA*

[5]*Department of Physics and Astronomy, University of Nebraska, Lincoln, Nebraska 68588, USA*

* Correspondence and requests for materials should be addressed to X.X. (xiaoshan.xu@unl.edu) or D.S. (email: dsun4@ncsu.edu) or J.S. (email: shenj5494@fudan.edu.cn).

**Abstract**

Manipulation of tunneling spin-polarized electrons via a ferroelectric interlayer sandwiched between two ferromagnetic electrodes, dubbed Multiferroic Tunnel Junctions (MFTJs), can be achieved not only by the magnetic alignments of two ferromagnets but also by the electric polarization of the ferroelectric interlayer, providing great opportunities for next-generation multi-state memory devices. Here we show that a $La_{0.67}Sr_{0.33}MnO_3$ (LSMO)/$PbZr_{0.2}Ti_{0.8}O_3$(PZT)/Co structured MFTJ device can exhibit multilevel resistance states in the presence of gradually reversed ferroelectric domains via tunneling electro-resistance and tunneling magnetoresistance, respectively. The nonvolatile ferroelectric control in the MFTJ can be attributed to separate contributions arising from two independent ferroelectric channels in the PZT interlayer with opposite polarization. Our study shows the dominant role of 'mixed' ferroelectric states on achieving accumulative electrical modulation of multilevel resistance states in MFTJs, paving the way for multifunctional device applications.

# I. INTRODUCTION

Multiferroic Tunnel Junctions (MFTJs) consisting of two ferromagnetic (FM) electrodes separated by a thin ferroelectric (FE) insulating layer have garnered great interest for potential multi-functional applications in the field of spintronics as a result of their switchable electric polarization with high tunability [1]. By switching the electric polarization of the FE layer, the resistance of the MFTJ changes in correspondence with the electrostatic effect, strain effect, and interface effect, etc. [2]. This has been demonstrated through tunnel electro-resistance (TER), which can be described by the Brinkman model using the barrier height of the insulator and the screening length of the electrodes [3-5]. The change in resistance scales with the fraction of ferroelectric polarization domains, from which a parallel conduction model of ferroelectric poled up and down domains has been proposed [6]. A similar large TER effect has been reported in a semiconductor-based tunnel junction with an additional Schottky barrier on the semiconductor surface, demonstrating the electrical modulation of both the height and the width of the barrier using ferroelectricity [7, 8].

Owing to the adjacent FM electrodes attached to the FE material, the tunneling resistance of the MFTJ can be tuned by applying an external magnetic field, dubbed tunneling magnetoresistance (TMR) [9]. The pronounced magnetoelectric coupling (MEC) at the FM-FE interface enables us to actively manipulate the magnetic properties of the FM electrodes by switching the polarity of the FE interlayer, including the interfacial magnetism [10, 11], magnetic anisotropy [12, 13], spin polarization [14], magnetic domain size [15], and its phase transition [16, 17], etc. [1, 18-20]. This leads to the electrically modulated TMR response (e.g., polarity and magnitude) in MFTJs by switching the polarity of the FE interlayer.[21-23] By utilizing a ferroelectric material as the tunnel barrier layer, the MFTJs possesses an additional degree of freedom to modulate the spin transport and spin injection by electrical means, which is ideally desirable for chip integration. These magnetoelectric functionalities have been employed for multi-functional spintronic devices like four-state memories [24-26], making ferroelectric materials promising candidates for computing technologies.

Here we demonstrate that the electrical control of both tunneling electron- and magneto-resistance can be achieved in *one* MFTJ device (i.e., LSMO/PZT/Co) using ferroelectricity. The fabricated device exhibits multilevel nonvolatile resistance states by accumulatively tuning the area fraction of the ferroelectric up and down domains in the PZT interlayer. The correlated TER and TMR states at the 'mixed' FE state are calculated using a two-channel parallel resistance circuit. Our work presents promising ferroelectric control of spin-dependent electron tunneling resistance in MFTJs towards next-generation bio-inspired neuromorphic computing applications.

## II. EXPERIMENTAL METHODS

Pulsed Laser Deposition (PLD) is used to grow 30 nm LSMO and then 5 nm PZT films epitaxially on the surface of a $SrTiO_3$ (STO) substrate. The topography and ferroelectric properties of the LSMO/PZT films are characterized by a Veeco Dimension 3100 Piezo-response Force Microscope (PFM) at room temperature with atomically flat surfaces and switchable electric polarization [27]. Patterned 10 nm thick Co layers and ~10 nm Au capping layers are deposited on top of the LSMO/PZT films by thermal evaporation in a high vacuum chamber (base vacuum $<1\times10^{-6}$ Pa) using a shadow mask to fabricate junctions with areas of 200×250 μm$^2$. The magnetic properties of the LSMO films and the junctions are characterized by a Quantum Design Superconducting Quantum Interference Device (SQUID). A Keithley 2400 source meter and Physical Property Measurement System (PPMS, Quantum Design) are used for charge and spin transport measurements with controllable magnetic field and temperature. For electrical nonvolatile control of TER and TMR, a pre-set voltage bias ($V_{MAX}$) is programmed by ramping the voltage from zero to a certain level in order to initialize the accumulative ferroelectric state of the PZT interlayer in the presence of $x$% FE domains poled down ($0 \leq x \leq 100$). The TER and TMR effects are determined by measuring the resistance of the device at a small voltage bias, $V_{MR}$ ($|V_{MR}| \leq |V_{MAX}|$) as a function of electric field (R-V curves for TER effect) and magnetic field (R-H curves for TMR effect), respectively.

### III. RESULTS AND DISCUSSION

**A. TER characterization in MFTJs**

A schematic diagram of the MFTJs is shown in Fig. 1(a). LSMO and Co films are used as ferromagnetic electrodes for spin injection and detection. The magnetic hysteresis loops of the FM films are presented in Fig. 1(b), showing different coercive fields of ~40 Oe and ~500 Oe for the LSMO and Co, respectively. The parallel and antiparallel magnetic alignments of the FM electrodes in MFTJs is achieved by sweeping the magnetic field. The ferroelectric tunnel barrier comprises a 5 nm PZT film with a switchable FE polarization. The FE polarization direction for an as-grown PZT film is pointing up. The reversal of the FE domain in this ultrathin PZT layer (i.e., from up to down direction) has been demonstrated by applying a positive voltage bias through the PFM tip (> +0.8 V) [27-29], and is not expected to be affected by the top electrode [30], suggesting a similar FE switching behavior in the MFTJ configuration. The tunneling behavior of the MFTJ in the parallel magnetic configuration was characterized using current-voltage (I-V) curves having the PZT layer poled up ($V_{MAX}$ = +3.0 V) and down ($V_{MAX}$ = -3.0 V), respectively [Fig. 1(c)][31]. The non-linear behavior of both curves are consistent with the Brinkman model [3, 25, 32]: At a small voltage bias, the normalized differential conductance in the MFTJ [$G(V)/G(0), G = \frac{dI}{dV}$] can be described by a parabolical dependence of bias voltage (V) while neglecting high-order terms:

$$\frac{G(V)}{G(0)} = 1 - \frac{d\sqrt{2m_e e}\Delta\phi}{12\hbar\bar{\phi}^{\frac{3}{2}}}V + \frac{d^2 m_e e}{4\hbar^2 \bar{\phi}}V^2 \tag{1}$$

where $d$ is the barrier thickness ($d$=5 nm in this work), $\bar{\phi} = \frac{1}{2}(\phi_1 + \phi_2)$, i.e. the average barrier height at the LSMO/PZT ($\phi_1$) and the PZT/Co interface ($\phi_2$), $\Delta\phi = |\phi_2 - \phi_1|$ is the difference of asymmetric interfacial barriers. Figure 1(d) shows an example of $G(V)/G(0)$ from the I-V curve with PZT poled up (scatters) and the Brinkman model fitting curve (the blue plot) from Eq.(1) using $\overline{\phi_\uparrow} = \frac{1}{2}(\phi_{\uparrow 1} + \phi_{\uparrow 2}) = 0.3 \ eV$ and $\Delta\phi_\uparrow = |\phi_{\uparrow 2} - \Delta\phi_{\uparrow 1}| = 0.1 \ eV$. Thus, $\phi_{\uparrow 1} = 0.35 \ eV, \phi_{\uparrow 2} = 0.25 \ eV$ can be estimated. Similarly, for the PZT poled down we get $\overline{\phi_\downarrow} = \frac{1}{2}(\phi_{\downarrow 1} + \phi_{\downarrow 2}) = 0.5 \ eV, \Delta\phi_\downarrow = |\phi_{\downarrow 1} - \Delta\phi_{\downarrow 2}| = 0.2 \ eV$, and $\phi_{\downarrow 1} = 0.4 \ eV, \phi_{\downarrow 2} = 0.6 \ eV$ as illustrated by the schematic band structure in Fig. 1(e). The TER effect, defined as TER = $\frac{R_\downarrow}{R_\uparrow}$ (where $R_\uparrow$ and $R_\downarrow$ is the resistance of MFTJs when the PZT layer is fully poled up and down, respectively) can be expressed as [33, 34]:

$$\text{TER} \approx \exp\left(\frac{\sqrt{2m_e}}{\hbar}\frac{\Delta u}{\sqrt{u_0}}d\right) \tag{2}$$

where $d$ is the thickness of the PZT, $\Delta u = \overline{\phi_\downarrow} - \overline{\phi_\uparrow}$, i.e. the barrier height changes upon reversal of the FE polarization, and $u_0$ is the potential barrier height of PZT which can be determined as its average value of up and down states according to Wentzel-Kramers-Brillouin approximation, i.e. $u_0 = \frac{1}{2}(\overline{\phi_\downarrow} + \overline{\phi_\uparrow})$.

**B. TMR characterization in MFTJs and electrical modulation**

The resistance of the MFTJ can also be tuned by an external magnetic. Figure 2 shows the multilevel resistance states of a MFTJ tuned by an external magnetic and an electric field alternatively. The multi-state measurements are conducted by first ramping the bias voltage to $V_{MAX}$ to achieve one of the nonvolatile accumulated FE states [Fig.2(c)] and then the voltage bias is reduced to $V_{MR}$ ($|V_{MR}| \leqslant |V_{MAX}|$) to probe the magnetoresistance of the MFTJ at this programmed FE state. By gradually tuning $V_{MAX}$, the FE state of PZT can be accumulatively switched from the fully poled up state, to the partially poled down state, and then to the fully poled down state [27]. By sweeping the magnetic field [Fig.2(a)], TMR measurements are performed at each FE state [Fig. 2(d)]. The TMR value is calculated by:

$$\text{TMR} = \frac{R_{AP} - R_P}{R_P} \times 100\% \tag{3}$$

where $R_P$ and $R_{AP}$ are the resistances of the MFTJ device with parallel and antiparallel magnetic alignments of the FM electrodes.

- For $V_{MAX} \geqslant +3.0$ V and $V_{MAX} \leqslant -3.0$ V, the $R_P$ of the MFTJ remain unchanged by changing the $V_{MAX}$, indicating that the two fully poled states have been established. Remarkably, the polarity of the TMR response changes from negative to positive sign when the poling bias ($V_{MAX}$) switches the FE polarization states of PZT from the up to the down state, consistent with previous literature report [23].

- The change of $R_P$ in the -3.0 V < $V_{MAX}$ ≤ 0 V region corresponds to the formation of the 'mixed' FE polarization state with both poled up and down domains. TMR versus $V_{MR}$ curves are plotted under different $V_{MAX}$ in Fig. 2(e). A typical $V_{MR}$ dependence with a peak of TMR value around $V_{MR}$ = 0 V was observed.[35, 36] The bias voltage shifts the Fermi level of FM electrodes and their spin polarizations, and consequently modify the TMR values according to the Julliére model (TMR = $\frac{2P_{LSMO}P_{Co}}{1-P_{LSMO}P_{Co}}$) [37]. The quantitative dependence of TMR on $V_{MR}$ depends on the spin-resolved band structure of the electrodes [36].

- For 0 V ≤ $V_{MAX}$ < 3.0 V, the TMR is less sensitive to $V_{MAX}$ considering the measurement uncertainties, while for $V_{MAX}$ < 0 V, the sign of TMR changes dramatically from negative to positive [see Fig. 2(d)-(e)]. This can be understood as that the FE domain in as-grown PZT films are naturally poled up, along the same direction of positive electric field ($V_{MAX}$ > 0 V). A further increased positive $V_{MAX}$ does not change the FE polarization states significantly [27, 28] as well as the TER and TMR response. In contrast, for $V_{MAX}$ <0 V, by gradually tuning the voltage level of $V_{MAX}$, the intermediate polarization states of PZT containing both FE up and down domains will be formed. Consequently, both TER and TMR changes as a function of $V_{MAX}$ until the fully FE poled down state is reached.

**C. Model and discussion**

Figure 3 shows the both measured and simulated TER responses in MFTJ using a two-channel model upon the $V_{MAX}$ modulation. The TER response of MFTJs is derived from I-V characteristics, which strongly depends on $V_{MAX}$ [Fig. 3(b)]. Considering two individual channels corresponding to the FE up and down domains respectively, TER values can be defined as the ratio of device resistance with $x\%$ FE down domains ($R_x$) to the one with fully FE poled up domains ($R_\uparrow$), i.e., $TER(x) \equiv \frac{R_x}{R_\uparrow}$. Here the device resistance ($R_x$) at an intermediated FE polarization states is expressed by:

$$\frac{1}{R_x} = \frac{x\%}{R_\downarrow} + \frac{1-x\%}{R_\uparrow} \tag{4}$$

$$x\% = \frac{\frac{R_\downarrow}{R_\uparrow}(1-\frac{R_x}{R_\uparrow})}{(1-\frac{R_\downarrow}{R_\uparrow})\frac{R_x}{R_\uparrow}} \times 100\% \tag{5}$$

Taking the device resistance at $x\%=0$ and at $x\%=100\%$ as the $R_\uparrow$ and $R_\downarrow$, $x\%$ can be derived from the device resistance state, $R_x$ at each FE state, as shown in Fig. 3(c).

The fraction of switched FE down domains as a function of $V_{MAX}$ is plotted in Fig. 3(d). We found the FE domain starts to switch at $V_{MAX} \leq -0.5$ V, where $x\%$ changes dramatically from 0% to 100%. By fine-tuning the poling voltage $V_{MAX}$, the fraction of the FE down domain in the intermediate FE state can be electrically modulated, leading to the gradual evolution of the device resistance.

The same parallel resistor circuit model is also employed to interpret the evolution of TMR response by switching the FE polarization of the PZT layer. Because of the antiparallel and parallel magnetic alignments of the two FM electrodes, there are two magnetic configurations at each FE state. Fig. 4(a-c) shows the schematic magnetic configurations in MFTJs for the FE fully poled up, partially poled down, and fully poled down states, respectively. Taking $V_{MR} = 0.2$ V and -0.25 V as an example, the $R(H)$ curves at two FE fully poled states (i.e., $x\% = 0$ and $x\% = 100\%$) are presented in Fig. 4(d-e). The shape of the $R(H)$ hysteresis changes significantly when the PZT polarization is reversed, suggesting a change of the magnetic anisotropy of the Co layer or even the existence of exchange bias. This is probably caused by the migration of oxygen vacancy and formation of CoO at the PZT/Co interface when the PZT is poled down [38].

Taking the FE state into account, the TMR value of the MFTJ can be described as:

$$\text{TMR} = \frac{R_{x,\text{AP}} - R_{x,\text{P}}}{R_{x,\text{P}}} = \frac{R_{\downarrow,\text{AP}} \cdot R_{\uparrow,\text{AP}}}{R_{\downarrow,\text{P}} \cdot R_{\uparrow,\text{P}}} \cdot \frac{x\% \cdot R_{\uparrow,\text{P}} + (1-x\%)R_{\downarrow,\text{P}}}{x\% \cdot R_{\uparrow,\text{AP}} + (1-x\%)R_{\downarrow,\text{AP}}} - 1 \qquad (6)$$

i.e., the TMR of intermediate FE states [Fig. 4(b)] can be estimated from the values of the two fully pole FE states [Fig. 4(a, c)]. The calculated results from Eq. (6) are presented in Fig. 4(f) for $V_{\text{MR}} = 0.2$ V and -0.25 V, which agree well with the measured values, suggesting the validity of the two-channel model.

## IV. CONCLUSIONS

LSMO/PZT/Co structured multiferroic tunnel junctions have been fabricated, and the transport of spin polarized charge carriers through the MFTJ via tunneling have been successfully modulated using the electric-field controllable ferroelectric states of the PZT layer. By fine-tuning the poling voltage, $V_{\text{MAX}}$, intermediate FE states have been established, leading to the gradual change of the TER and the TMR responses of the device with multilevel resistance states, where the electric field and the magnetic field serve as coarse and fine adjustments respectively. Our work suggests a feasible approach to electrically tune the nonvolatile states of the FE layer, where both the TER and the TMR responses can be derived using a two-channel model corresponding to two FE domains. The electrically modulated intermediate FE polarization states in MFTJs can be used for multi-state memories applications.

## ACKNOWLEDGMENTS

This work was supported by the National Key Research and Development Program of China (2017YFB0305500), National Natural Science Foundation of China (11504055), the Hunan Provincial Natural Science Foundation of China (2018JJ2480) and Fundamental Research Funds of Central South University (2018zzts324). E.V. and D.S. were thankful for the start-up support provided by North Carolina State University and NC State-Nagoya Collaboration Grant. The sample synthesis work at ORNL was supported by the U.S. Department of Energy, Office of Science, Basic Energy Sciences, Materials Sciences and Engineering Division.

**Figures and captions**

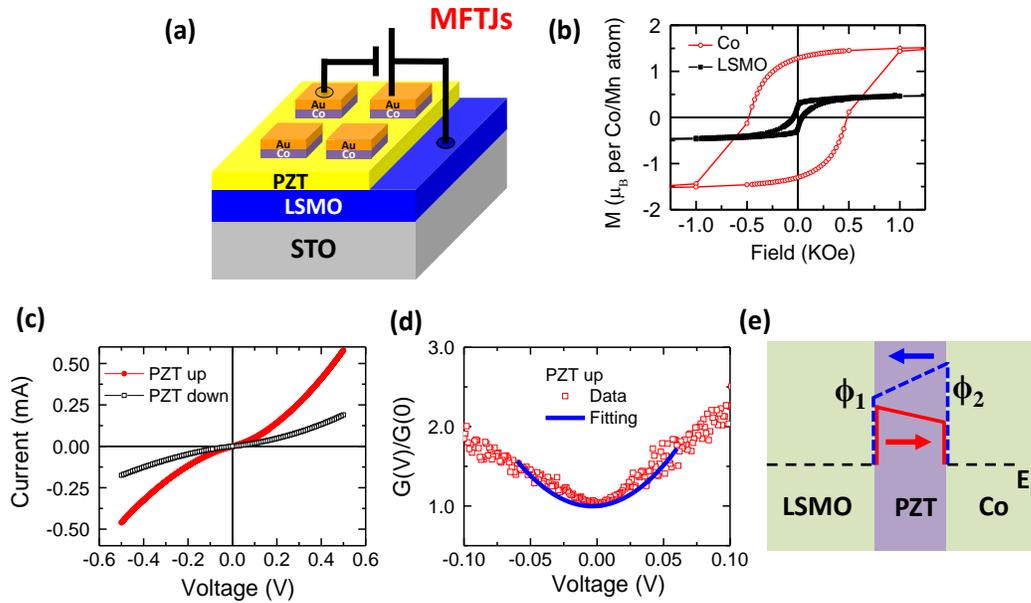

**FIG. 1. Multiferroic tunnel junctions.** (a) Schematic image of the LSMO/PZT/Co MFTJs. (b) Magnetic hysteresis loops of the LSMO and Co films used from FM electrodes with coercive fields of 40 Oe and 500 Oe, respectively. (c) IV curves of the junction (measured at 10 K and 2000 Oe) with the PZT poled up and down, respectively. (d) Dynamic conductance (red square dots, PZT poled up) as a function of voltage, calculated from the I-V curve and the fit by the Brinkman model (blue solid line) in small voltage range. (e) Energy diagram of the MFTJ with PZT poled up and down, respectively.

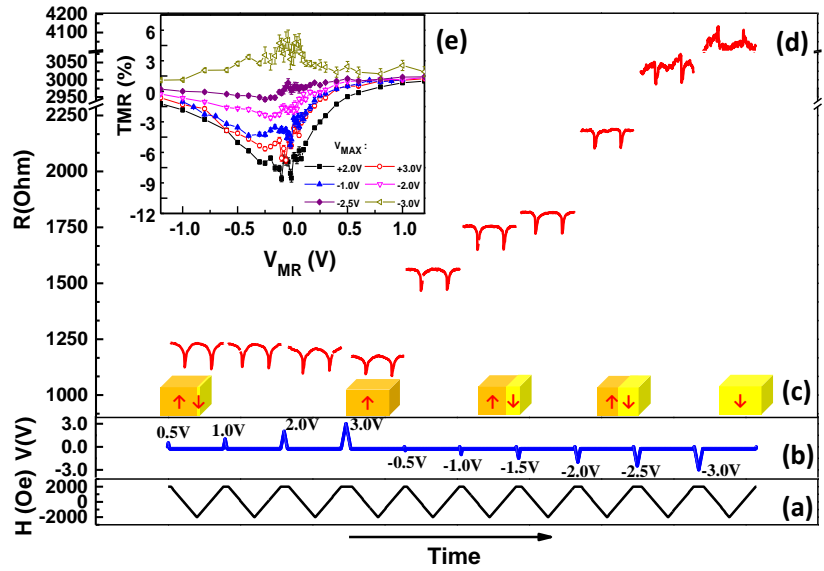

**FIG. 2. Multilevel resistance states of the MFTJ tuned by external magnetic and electric field.** (a-d) Magnetic field and electric field programing of resistance: the voltage was ramped to $V_{MAX}$ = 0.5 V, 1.0 V, 2.0 V, 3.0 V, -0.5 V, -1.0 V, -1.5 V, -2.0 V, -2.5 V, -3.0 V (b) to modulate the FE states of the PZT layer (c), and then to $V_{MR}$ (= -0.25 V) for magnetoresistance measurements (d) by sweeping the magnetic field (a). (e) TMR as a function of $V_{MR}$ measured after application of different $V_{MAX}$.

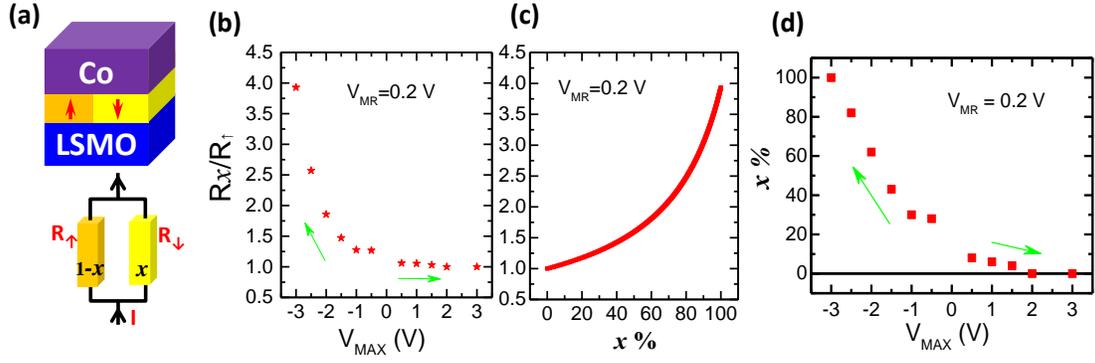

**FIG. 3. Modulation and simulation of the TER of the MFTJ with intermediate ferroelectric polarization state ($x\%$).** (a) Schematics of the two-channel model consisting of up and down ferroelectric domains which correspond to different resistivities. (b) TER defined as the resistance ratio $\frac{R_x}{R_\uparrow}$ as a function of the poling voltage ($V_{MAX}$). (c) TER as a function of the fraction of FE down domains ($x\%$) calculated from the two-channel model and the maximum and minimum resistance values (see text). (d) The $x$ value as a function of $V_{MAX}$, extracted according to (b) and (c). The arrows in (b) and (d) indicate measurement sequence.

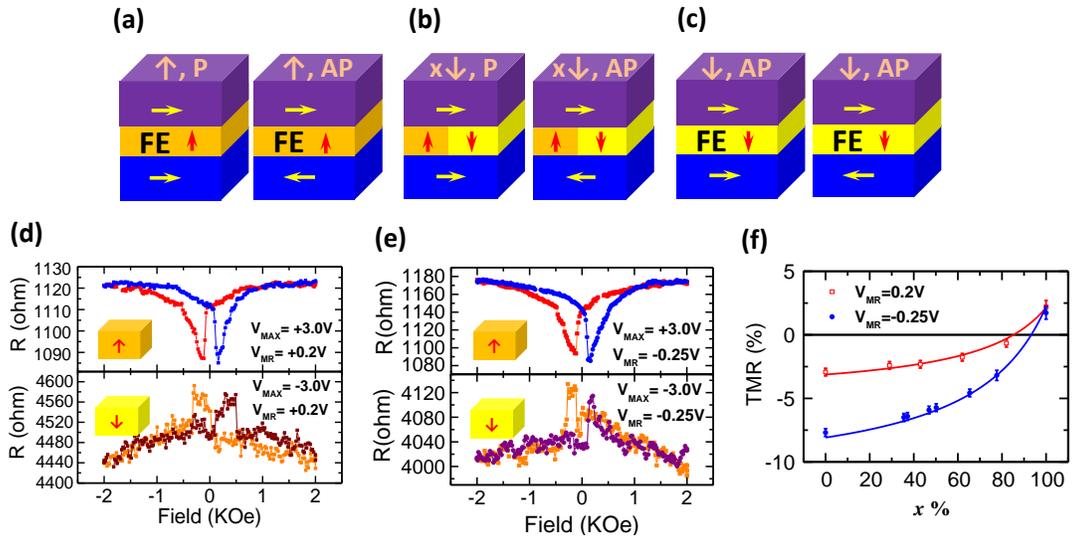

**FIG. 4. Modulation and simulation of the TMR of the MFTJ with intermediate ferroelectric polarization state ($x\%$).** (a-c) Schematics of the MFTJ with the magnetization of the ferromagnetic electrodes parallelly or antiparallelly aligned and with the ferroelectric layer fully poled up ($x\% = 0$), partly poled down ($x\%$), or fully poled down ($x\% = 100\%$). (d) $R(H)$ of the MFTJ measured at $V_{MR} = 0.2$ V and (e) $V_{MR} = -0.25$ V for the $x\% = 0$ [(a) state] and the $x\% = 100\%$ [(c) state]. (f) Measured (scatters) TMR compared with the TMR calculated (lines) from the two-channel model for $V_{MR} = 0.2$ V and $V_{MR} = -0.25$ V, respectively.